\documentclass[aps,jcp,twocolumn,superscriptaddress,floatfix]{revtex4}

\usepackage{dcolumn} 
\usepackage{bm} 
\usepackage{multirow} 
\usepackage{pifont} 
\usepackage{epsfig}
\usepackage{fixltx2e} 
\usepackage{amsmath} 
\usepackage{physics} 
\begin{document}

\author{Run R. Li}
\affiliation{
             Department of Chemistry and Biochemistry,
             Florida State University,
             Tallahassee, FL 32306-4390}
\author{A. Eugene DePrince III}
\affiliation{
             Department of Chemistry and Biochemistry,
             Florida State University,
             Tallahassee, FL 32306-4390}
\email{deprince@chem.fsu.edu}

\title{The role of orbital angular momentum constraints in the variational
optimization of the two-electron reduced-density matrix}



\begin{abstract}

The direct variational determination of the two-electron reduced-density
matrix (2-RDM) is usually carried out under the assumption that the 2-RDM
is a real-valued quantity.  However, in systems that possess orbital
angular momentum symmetry, the description of states with a well-defined,
non-zero $z$-projection of the orbital angular momentum requires a
complex-valued 2-RDM.  We consider a semidefinite program
suitable for the direct optimization of a complex-valued 2-RDM and explore
the role of orbital angular momentum constraints in systems that possess
the relevant symmetries. For atomic systems, constraints on the
expectation values of the square and $z$-projection of the orbital angular
momentum operator allow one to optimize 2-RDMs for multiple orbital
angular momentum states.  Similarly, in linear molecules, orbital angular
momentum projection constraints enable the description of multiple
electronic states, and, moreover, for states with a non-zero
$z$-projection of the orbital angular momentum, the use of complex-valued
quantities is essential for a qualitatively correct description of the
electronic structure.  For example, in the case of molecular oxygen, we
demonstrate that orbital angular momentum constraints are necessary to
recover the correct energy ordering of the lowest-energy singlet and
triplet states near the equilibrium geometry.  However, care must still be
taken in the description of the dissociation limit, as the 2-RDM-based
approach is not size consistent, and the size-consistency error varies
dramatically, depending on the $z$-projections of the spin and orbital
angular momenta.

\end{abstract}

\maketitle

\section{Introduction}

It has long been understood that the direct variational determination of
the elements of the two-electron reduced-density matrix (2-RDM) is a
desirable prospect.\cite{Husimi:1940:22,Lowdin:1955:1474,Mayer:1955:100}
The 2-RDM affords a much more compact representation of the electronic
structure than is offered by the $N$-electron wavefunction, and, yet, it
contains sufficient information to exactly specify the electronic energy
for any many-electron system.  Hence, the wavefunction can, in principle,
be supplanted by the 2-RDM in variational calculations, provided that the
space of 2-RDMs over which the optimization is performed is restricted to
contain only those that derive from antisymmetrized $N$-electron
wavefunctions.  Such 2-RDMs are said to be
$N$-representable.\cite{Coleman:1963:668}  One of the strengths of
2-RDM-based methods is that they are naturally multiconfigurational and
can thus be applied to multireference or strongly-correlated electronic
structure problems.  Indeed, variational 2-RDM (v2RDM)
approaches\cite{Garrod:1975:868,Mihailovic:1975:221,Rosina:1975:300,Erdahl:1979:1366,Erdahl:1979:147,Nakata:2001:8282,Mazziotti:2001:042113,Mazziotti:2002:062511,Mazziotti:2006:032501,Zhao:2004:2095,Fukuda:2007:553,Cances:2006:064101,Verstichel:2009:032508,FossoTande:2016:423,Verstichel:2011:1235}
that enforce necessary ensemble $N$-representability
conditions\cite{Garrod:1964:1756,Zhao:2004:2095,Erdahl:1978:697} can be
used to realize a polynomially-scaling
approximation\cite{Gidofalvi:2008:134108,FossoTande:2016:2260} to complete
active space self-consistent field (CASSCF)
theory\cite{Roos:1980:157,Siegbahn:1980:323,Siegbahn:1981:2384,Roos:1987:399}
that is applicable to active spaces composed of as many as 64 electrons
in 64 orbitals.\cite{Mullinax:2019:submitted}

Such nice properties notwithstanding, v2RDM approaches suffer from a
number of well-known issues that limit their application to general
quantum chemical problems.  For example, the methods sometimes dissociate
molecules into fractionally charged
species.\cite{vanAggelen:2009:5558,Verstichel:2010:114113,vanAggelen:2011:054115}
The source of this error is the lack of a derivative discontinuity in the
energy when considering fractionally charged atoms; the same issue arises
within density functional theory.\cite{Cohen:2008:792}  Second, the direct
application of the v2RDM approach to excited states is an outstanding
problem.  Spin-symmetry constraints give one access to multiple
(lowest-energy) spin states, but, even then, one cannot reliably compare
states that have the same total spin angular momentum but different 
$z$-projections,
as known $N$-representability conditions do not
constrain the 2-RDMs representing these states
equally.\cite{vanAggelen:2012:014110}. The next logical step would be the
application of spatial symmetry constraints to
differentiate electronic states. However, this strategy cannot be easily 
realized 
within the v2RDM framework because
the point-group of the molecule is an $N$-electron property, the
evaluation of which requires knowledge of the $N$-electron reduced-density
matrix.

This work aims to at least partially address this last deficiency of the
v2RDM approach.  In systems possessing well-defined orbital angular
momentum symmetry (i.e., atoms and linear molecules), the application of
appropriate orbital angular momentum constraints allows for the direct
description of multiple electronic states with different spatial
symmetries.  The application of v2RDM techniques to atomic states with
non-zero magnitude and $z$-projection of the orbital angular momentum
requires the consideration of complex-valued reduced-density matrices
(RDMs).  While atomic states with non-zero magnitude and zero
$z$-projection of the orbital angular momentum can be described with
real-valued RDMs, we show that the quality of the energy is inferior to
that corresponding to non-zero $z$-projection states.  This behavior is
reminiscent of that observed for different spin angular momentum
projection states in Ref. \citenum{vanAggelen:2012:014110}.  For linear
molecular systems, we demonstrate that angular momentum constraints and
complex RDMs can be necessary for even a qualitatively correct description
of the electronic structure; for example, in a correlation-consistent
polarized valence double-zeta (cc-pVDZ)\cite{Dunning:1989:1007} basis set,
a real-valued v2RDM computation incorrectly predicts that the
lowest-energy state of molecular oxygen is a singlet.

This paper is organized as follows.  Section \ref{SEC:THEORY} outlines the
general procedure for the direct determination of the 2-RDM under ensemble
$N$-representability conditions and describes how one can incorporate
orbital angular momentum constraints into the optimization.  Section
\ref{SEC:COMPUTATIONAL_DETAILS} then provides some of the technical
details of our computations.  We explore the role of orbital angular
momentum constraints in atomic and linear molecular systems in Sec.
\ref{SEC:RESULTS}, and some concluding remarks are provided in Sec.
\ref{SEC:CONCLUSIONS}.

\section{Theory}

\label{SEC:THEORY}

\subsection{The variational optimization of the 2-RDM}

The electronic energy of a many-electron system is a linear functional of
the one-electron reduced-density matrix (1-RDM) and the 2-RDM:
\small
\begin{align}
	\label{EQN:energy}
        E & =   \frac{1}{2} \sum_{pqrs} ~({}^2D^{p_{\alpha}q_{\alpha}}_{r_{\alpha}s_{\alpha}} 
            + {}^2D^{p_{\alpha}q_{\beta}}_{r_{\alpha}s_{\beta}} 
            + {}^2D^{p_{\beta}q_{\alpha}}_{r_{\beta}s_{\alpha}} 
            + {}^2D^{p_{\beta}q_{\beta}}_{r_{\beta}s_{\beta}} ) ( pr | qs )  \nonumber \\ 
        &  +  \sum_{pq} ~( {}^1D^{p_{\alpha}}_{q_\alpha} + {}^1D^{p_{\beta}}_{q_\beta} )  h_{pq}.
\end{align}
\normalsize
Here, $(pr|qs)$ represents a two-electron repulsion integral, $h_{pq}$
represents the sum of the one-electron kinetic energy and electron/nuclear
potential energy integrals, and the summation indices run over all spatial
orbitals.  The 1-RDM and 2-RDM can be expressed in second-quantized
notation as
\begin{equation}
	{}^1D^{p_{\sigma}}_{q_{\sigma}} = \langle \Psi | \hat{a}^\dagger_{p_{\sigma}}\hat{a}_{q_{\sigma}} | \Psi \rangle,
\end{equation}
and
\begin{equation}
	{}^2D^{p_{\sigma}q_{\tau}}_{r_{\sigma}s_{\tau}} = \langle \Psi | \hat{a}^\dagger_{p_{\sigma}}\hat{a}^\dagger_{q_{\tau}} \hat{a}_{s_{\tau}}\hat{a}_{r_{\sigma}}| \Psi \rangle,
\end{equation}
respectively,  where $\hat{a}^\dagger$ ($\hat{a}$) represents a fermionic
creation (annihilation) operator, and, throughout,  Greek labels represent
either $\alpha$ or $\beta$ spin.  The 1- and 2-RDM can be determined
directly via the minimization of Eq. \ref{EQN:energy} with respect to
variations in their elements, provided that the optimization is
constrained such that it considers only those reduced-density matrices
(RDMs) that are derivable from an ensemble of antisymmetrized $N$-electron
wavefunctions.  In practical computations, we can only reasonably enforce
{\em approximate} $N$-representability conditions, and the resulting
energy is thus a lower-bound to the exact (full configuration interaction
[CI]) energy within the relevant basis set.  In this work, we consider the
two-particle (``PQG'') $N$-representability constraints of Garrod and
Percus.\cite{Garrod:1964:1756}

As we are concerned with non-relativistic Hamiltonians, we also enforce
constraints on the spin structure of the 1- and 2-RDM.  For example, the
total spin of the system is related to an off-diagonal trace of the 2-RDM,\cite{Perez:1997:55,Gidofalvi:2005:052505}
\begin{equation}
        \label{EQN:s2}
        \sum_{pq}{}^2D^{p_{\alpha}q_{\beta}}_{q_{\alpha}p_{\beta}} = \frac{1}{2} ( N_\alpha + N_\beta ) + M_S^2 - S(S+1),
\end{equation}
where $S$ and $M_S$ represent the total spin and spin-projection quantum
numbers, respectively. In addition, in all computations presented herein,
the RDMs are constrained to represent maximal spin-projection states, as
it has been demonstrated that such states are better described by v2RDM
methods than other spin-projection states.\cite{vanAggelen:2012:014110}
Maximal spin-projection states must satisfy
\begin{equation}
\label{EQN:raising}
\hat{S}^+|\Psi\rangle=0, 
\end{equation}
where $\hat{S}^+$ represents a spin angular momentum raising operator.
Eq. \ref{EQN:raising} implies a weaker set set of constraints of the
form\cite{vanAggelen:2012:014110}
\begin{equation}
\label{EQN:raising_weaker}
\forall r_{\beta}, s_{\alpha}: \langle \Psi | \hat{a}_{r_{\beta}}^\dagger\hat{a}_{s_{\alpha}}\hat{S}^+|\Psi\rangle=0, 
\end{equation}
which can be expressed in terms of the one-particle one-hole RDM
(${}^2{\bf G}$)
\begin{equation}
        \label{EQN:S+_1}
        \forall r_{\beta}, s_{\alpha}: \sum_p {}^2G_{p_{\beta}p_{\alpha}}^{r_{\beta}s_{\alpha}}=0,
\end{equation}
whose elements are given by
\begin{equation}
        \label{EQN:G2_1}
	{}^2G^{p_{\sigma}q_{\tau}}_{r_{\lambda}s_{\mu}} = \langle \Psi | \hat{a}^\dagger_{p_{\sigma}}\hat{a}_{q_{\tau}} \hat{a}^\dagger_{s_{\mu}}\hat{a}_{r_{\lambda}}| \Psi \rangle.
\end{equation}
Similarly, the adjoint of the raising operator acting on the bra space
also annihilates the state, giving rise to a complementary set 
of constraints
\begin{equation}
        \label{EQN:S+_2}
        \forall r_{\beta}, s_{\alpha}: \sum_p {}^2G_{r_{\beta}s_{\alpha}}^{p_{\beta}p_{\alpha}}=0.
\end{equation}

The direct variational optimization of the 1- and 2-RDM subject to the
constraints outlined above constitutes a semidefinite programming (SDP)
problem.  We solve this problem using a modified boundary-point SDP
algorithm\cite{Povh:2006:277,Malick:2009:336, Mazziotti:2011:083001}
similar to that described in Ref. \citenum{FossoTande:2016:2260}.  As
discussed below, the introduction of orbital angular momentum constraints
requires that the boundary-point algorithm be generalized to treat complex
RDMs.

\subsection {Orbital angular momentum constraints}

\label{SEC:THEORY_OAM}

Consider the Hamiltonian for an atomic many-electron system.  At the
non-relativistic limit, the operators corresponding to the square of the
orbital angular momentum ($\hat{L}^2$) and its projection onto the
$z$-axis ($\hat{L}_z$) commute with this Hamiltonian.  Hence, RDMs
corresponding to good orbital angular momentum states should satisfy
additional equality constraints, including
\begin{equation}
	\label{EQN:L2}
	\langle \Psi | \hat{L}^2 | \Psi \rangle = L(L+1),
\end{equation}
and
\begin{equation}
	\label{EQN:Lz}
	\langle \Psi | \hat{L}_z | \Psi \rangle = M_L,
\end{equation}
where $L$ and $M_L$ represent the total orbital angular momentum and
orbital angular momentum projection quantum numbers, respectively.  These
constraints can be expressed in terms of the elements of the 1- and 2-RDM
as
\begin{eqnarray}
	\label{EQN:L2RDM}
    \sum_{\xi=x,y,z} &\bigg (& \sum_{\sigma \tau}\sum_{pqrs} {}^2D^{p_{\sigma}r_{\tau}}_{q_{\sigma}s_{\tau}} [L_\xi]^p_q [L_\xi]^r_s \nonumber \\
    &+& \sum_{\sigma} \sum_{pq} {}^1D^{p_{\sigma}}_{q_{\sigma}} [L_\xi^2]^p_q \bigg )  
    =  L(L+1),
\end{eqnarray}
and
\begin{equation}
	\label{EQN:LzRDM}
	\sum_{\sigma}\sum_{pq} {}^1D^{p_{\sigma}}_{q_{\sigma}} [L_z]^p_q = M_L,
\end{equation}
where $[L_\xi]^p_q$ and $[L_\xi^2]^p_q$ represent matrix elements of the
$\xi$-component of the angular momentum operator and its square,
respectively.

A 1-RDM that satisfies Eq. \ref{EQN:LzRDM} is not guaranteed to represent
a wavefunction that is an eigenfunction of $\hat{L}_z$.  Accordingly, we
also consider a constraint on the variance in $\hat{L}_z$, $(\Delta L_z)^2
= \langle \hat{L}_z^2 \rangle - \langle \hat{L}_z \rangle ^2$, which can
be evaluated with knowledge of the 2-RDM as
\begin{eqnarray}
    (\Delta L_z)^2 &=& \sum_{\sigma\tau}\sum_{pqrs} {}^2D^{p_{\sigma}r_{\tau}}_{q_{\sigma}s_{\tau}} [L_z]^p_q [L_z]^r_s \nonumber \\
                   &+& \sum_{\sigma} \sum_{pq} {}^1D^{p_{\sigma}}_{q_{\sigma}} [L_z^2]^p_q - M_L^2.
\end{eqnarray}
Here, we have assumed that the 1-RDM satisfies Eq. \ref{EQN:LzRDM}, and,
thus, $\langle \hat{L}_z \rangle ^2 = M_L^2$.  Similar arguments could be
made for RDMs that satisfy Eq.  \ref{EQN:L2RDM}, so a constraint on the
variance of $\hat{L}^2$, $(\Delta L^2)^2 = \langle \hat{L}^4 \rangle -
\langle \hat{L}^2 \rangle ^2$, might also be desirable.  However, the
evaluation of this quantity requires knowledge of the four-particle RDM,
so this constraint will not be considered in this work.

Since the angular momentum operator is pure imaginary, the RDMs that enter
our computations can only represent states with non-zero $M_L$ if they are
allowed to take on complex values.  Although the boundary-point SDP
algorithm was initially defined using real matrices, its extension to the
optimization of complex and even quaternion matrices is a purely technical
challenge.  \cite{Goemans:2004:442,Wolkowicz:2012:book} Realizing that the
field of complex matrices, ${\bf M}$, is isomorphic to the field of $2
\times 2$ real matrices of the form
\begin{equation}
\label{EQN:isomorphism}
\Re({\bf M}) + i \Im({\bf M}) \simeq
\left[
	\begin{array}{cc}
		\Re({\bf M}) & -\Im({\bf M}) \\
		\Im({\bf M}) &  \Re({\bf M}) 
	\end{array}
\right], 
\end{equation}
one can map the complex SDP programming problem to a real one with RDMs of
twice the original dimension, and, thus, a conventional SDP algorithm can
be applied.

As discussed in Refs. \citenum{Mazziotti:2011:083001} and
\citenum{FossoTande:2016:2260}, the boundary-point SDP solver for the
v2RDM problem is a two-step procedure.  In the first step, the dual
solution to the SDP ({\bf y}) is updated by solving
\begin{equation}
\label{EQN:CG}
{\bf A}{\bf A}^T{\bf y} = {\bf A}({\bf c} - {\bf z}) + t ({\bf b} - {\bf A}{\bf x})
\end{equation}
using conjugate gradient (CG) techniques.  Here, {\bf x} represents the
primal solution vector (which maps onto the RDMs), {\bf y} and {\bf z}
represent dual solution vectors, {\bf c} represents a vector containing
the one- and two-electron integrals that define the quantum system, and
{\bf A} and {\bf b} represent the constraint matrix and vector,
respectively, which encode the $N$-representability conditions.  The
symbol $t$ represents a penalty parameter.  In the second step, the primal
solution {\bf x} and the secondary dual solution {\bf z} are updated via
the solution of an eigenvalue problem.  The rate-limiting step in this
algorithm is the latter one, and its computational cost increases with the
third-power of the dimension of the RDMs. As such, expanding the complex
RDMs as is done in Eq. \ref{EQN:isomorphism} will increase the number
of floating-point operations required by the boundary-point SDP algorithm
by a factor of eight.

We have performed numerical tests to determine the relative efficiency of
real symmetric (\texttt{DSYEV}) and complex Hermitian (\texttt{ZHEEV})
eigensolvers.  The wall time required to diagonalize a complex matrix of
dimension 4000 is roughly 30\% of that required for the diagonalization of
a real symmetric matrix of twice the dimension, when using Intel's MKL
library and one core of an Intel Core i7-6850K CPU.  Hence, we elect to
retain the use of complex RDMs and modify the boundary-point solver
accordingly.  The only substantive change is that the number of coupled
linear equations represented by Eq. \ref{EQN:CG} increases by a factor of
two; one set of equations is used to update $\Re({\bf y})$, while the
other determines $\Im({\bf y})$.  Because the constraints we consider do
not directly couple the real and imaginary components of the RDMs, these
equations can be solved independently.

\section{Computational details}

\label{SEC:COMPUTATIONAL_DETAILS}

The boundary-point SDP solver for the complex v2RDM problem was
implemented as a plugin to the \textsc{Psi4} electronic structure
package.\cite{Psi4:1:1} Optimized RDMs obtained from this plugin satisfied
the PQG $N$-representability conditions and the spin angular momentum
constraints outlined in Sec.  \ref{SEC:THEORY}.  Energies from v2RDM
computations were compared to those from full CI and multireference CI
(MRCISD+Q) computations performed with the \textsc{Psi4} and ORCA
\cite{Neese:2018:e1327} packages, respectively.  All orbitals were
considered active within all v2RDM and full CI computations, while the
reference computations for MRCISD+Q considered only full valence active
spaces.  All computations on atomic systems employed the cc-pVDZ basis
set, while linear molecular systems were described by the
STO-3G\cite{STO3G}, Dunning-Hay double zeta (D95V)\cite{DunningHay},
6-31G*,\cite{Hehre:1972:2257,Hariharan:1973:213,Francl:1982:3654} and
cc-pVDZ basis sets; the reader is referred to Sec.
\ref{SEC:RESULTS_LINEAR} for additional details.

For atomic systems, the v2RDM
procedure was considered converged when $\epsilon_{\rm error} < 1.0 \times
10^{-5}$ and $\epsilon_{\rm gap} < 1.0 \times 10^{-4}\;E_{\rm h}$, with
the exception of two cases identified in Table
\ref{TAB:LSTATES_ExcitationE} for which the convergence were achieved
at least at $\epsilon_{\rm error} < 4.4 \times 10^{-6}$ and 
$\epsilon_{\rm gap} < 5.6 \times 10^{-4}\;E_{\rm h}$. Here, $\epsilon_{\rm error}$
refers to the maximum of the primal error ($|| {\bf A x} - {\bf b} ||$)
and dual error ($|| {{\bf A}^T {\bf y}-{\bf c}+{\bf z}}  ||$), and the
primal/dual energy gap, $\epsilon_{\rm gap}$,  is defined as $| {\bf
x}^T{\bf c} - {\bf b}^{T}{\bf y}|$.  
For linear molecular systems, the v2RDM procedure was considered converged
when $\epsilon_{\rm error} < 1.0 \times 10^{-4}$ and $\epsilon_{\rm gap} <
1.0 \times 10^{-4}\;E_{\rm h}$, with the exception of several 
calculations used to produce Fig. \ref{FIG:O2_ENERGY_DIAGRAM}. The 
most challenging calculation could only be converged to $\epsilon_{\rm error} < 1.4 \times 10^{-5}$ and
$\epsilon_{\rm gap} < 2.0 \times 10^{-3}\;E_{\rm h}$, and six other
calculations were converged to at least $\epsilon_{\rm error} < 1.2 \times
10^{-5}$ and $\epsilon_{\rm gap} < 8.3 \times 10^{-4}\;E_{\rm h}$. 
The reader is referred to the Supporting Information for additional details.

All v2RDM computations exploited the block structure of the RDMs resulting
from spin and abelian point-group symmetry considerations, but it should
be noted that the point group was chosen in each case such all operators
belonged to the totally symmetric irreducible representation.  Hence,
computations in which we constrained the expectation values of $\hat{L}_z$
were performed within the $C_{2h}$ point group, and computations in which
we constrained the expectation value of $\hat{L}^2$ were performed within
the $C_i$ point group.

The orbital angular momentum constraints outlined in Sec.
\ref{SEC:THEORY_OAM} involve molecular integrals that do not usually arise
in quantum chemical energy calculations.  The molecular integrals over the
orbital angular momentum operator, $\hat{L}_z$, were
obtained from the standard molecular integral library in \textsc{Psi4}.
On the other hand, the integrals over the square of the angular momentum
operator are not implemented in this package.  We evaluated integrals of
the form $[L_{\xi}^2]^p_q = \langle \chi_p| \hat{L}^2_{\xi} |\chi_q
\rangle$ numerically, where $\xi\in \{x,y,z\}$, and $\chi_p$ represents an
atomic basis function.  Numerical integrals were evaluated on the same
quadrature grids employed with density functional theory (DFT)
computations in \textsc{Psi4}.  We use the Lebedev-Trueutler (75,302)
grid, which is the default grid for all DFT computations in \textsc{Psi4}.

\section{Results and discussion}

\label{SEC:RESULTS}
\begin{table}

    \caption{Designation of the v2RDM computations on atomic systems
    according to the complexity of the RDMs and the orbital angular
    momentum constraints enforced.}

    \label{TAB:LABELS} 

    \centering
    \begin{tabular}{lcccl}
        \hline
        \hline
designation       & ~~~ & RDM complexity & ~~~ & constraints enforced \\
\hline
real              & ~~~ & real           & ~~~ &                         \\
complex           & ~~~ & complex        & ~~~ &                         \\
L$^2$             & ~~~ & complex        & ~~~ & ~~$\langle \hat{L}^2 \rangle$             \\
L$_z$             & ~~~ & complex        & ~~~ & ~~$\langle \hat{L}^2 \rangle$, $\langle \hat{L}_z \rangle $ \\
$(\Delta$L$_z)^2$ & ~~~ & complex        & ~~~ & ~~$\langle \hat{L}^2 \rangle$, $\langle \hat{L}_z \rangle $, $(\Delta L_z)^2$ \\
                \hline
                \hline
    \end{tabular}
\end{table}

In this Section, we numerically evaluate the effects of orbital angular
momentum constraints in v2RDM computations on systems with well-defined
orbital angular momentum symmetry.  Table \ref{TAB:LABELS} provides the
designations used to describe the constraints applied in calculations on
atomic systems, as well as the complexity of the RDMs.  Note that the
consideration of $\hat{L}^2$ symmetry does not require the use of complex
RDMs, but L$^2$ computations were performed using our complex-valued v2RDM
algorithm nonetheless.  

\subsection{Atomic systems}
\label{SEC:RESULTS_ATOMIC}

\begin{figure}[!htpb]

    \caption{Errors in ground-state energies (mE$_{\rm h}$) of second row
    atoms computed at the v2RDM/cc-pVDZ level of theory, as compared to
    results from full CI.}

    \label{FIG:GROUNDSTATES}
    \begin{center}
        \includegraphics[scale=1.0]{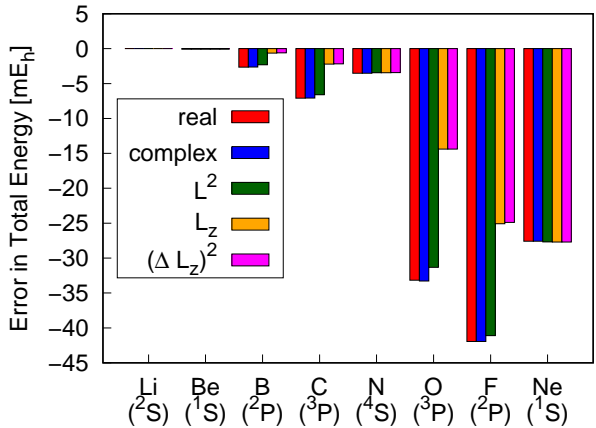}
    \end{center}

\end{figure}

Figure \ref{FIG:GROUNDSTATES} illustrates the errors in the ground-state
energies of second-row atoms computed at the v2RDM level of theory,
relative to energies obtained from full CI computations.  First, as a
technical note, the error incurred when using complex- and real-valued
RDMs is nearly indistinguishable on this scale, which suggests that our
complex-valued boundary-point SDP algorithm is implemented correctly.
Second, we note that the error increases, in general, with the number of
electrons.  This observation is consistent with the fact that v2RDM
methods with approximate $N$-representability constraints are not
strictly size extensive.  However, in the absence of orbital angular
momentum constraints, the error does not increase monotonically with
system size; it is exaggerated for states with non-zero orbital angular
momentum.   For these states, the application of $\hat{L}^2$ constraints
results in a minor improvement. On the other hand,
constraints on the expectation value of $\hat{L}_z$ lead to a significant
improvement in accuracy.  Here, these non-zero angular momentum states
are taken to have the maximal orbital angular momentum, which results in
complex-valued RDMs.  The subsequent application of variance constraints
[$(\Delta L_z)^2 = 0$] leads to essentially no improvement in the
description of these maximal orbital angular momentum projection states.

Clearly, orbital angular momentum constraints play an important role in
the v2RDM-based description of ground states with non-zero total angular
momentum.  The data in Fig. \ref{FIG:GROUNDSTATES} indicate that, in some
cases (boron, carbon, and oxygen), the application of such constraints
reduces the error in the v2RDM energy by more than a factor of two.
Moreover, angular momentum constraints also allow us to directly
optimize 2-RDMs for excited states that are not otherwise accessible by
v2RDM methods.  Table \ref{TAB:LSTATES_ExcitationE} illustrates energy
differences between excited spin and orbital angular momentum states and
the ground electronic states for all second-row atoms, except lithium and
neon.  Note that all results tabulated under the heading ``L$_z$''
correspond to the maximum orbital angular momentum projection.  First, we
consider those states that are accessible without angular momentum
constraints (all cases in Table \ref{TAB:LSTATES_ExcitationE} for which
numerical values are given under the heading ``real'').  For the beryllium
atom, the \textsuperscript{1}S $\to$ \textsuperscript{3}P transition is
equally well-described by all combinations of angular momentum constraints
considered.  On the other hand, the description of every other transition
energy is improved by the consideration of angular momentum constraints,
sometimes dramatically so.  In particular, the consideration of
$\hat{L}^2$ symmetry improves the almost 1 eV error in the description of
the \textsuperscript{4}S $\to$ \textsuperscript{2}D transition in nitrogen
by 0.32 eV.  The subsequent application of the constraint on $\langle
\hat{L}_z \rangle$ reduces the error to only 0.15 eV.

Now, consider those cases in Table \ref{TAB:LSTATES_ExcitationE} where no
numerical values are given under the heading ``real;'' the excited states
in question are inaccessible to the v2RDM approach unless angular momentum
constraints are imposed.  In one case,  the \textsuperscript{4}S $\to$
\textsuperscript{4}P transition in nitrogen, a constraint on the
expectation value of $\hat{L}^2$ yields a terrible estimate of the
excitation energy; it is too low by 5.78 eV.  However, subsequent
application of the constraint on $\langle \hat{L}_z \rangle$ yields an
excitation energy that agrees with that from the full CI to within less
than 0.01 eV.  We also observe that the application of the $\hat{L}_z$
constraint improves over the consideration of the $\hat{L}^2$ constraint
alone for the \textsuperscript{4}S $\to$ \textsuperscript{2}P transition
in nitrogen, although the improvement is less dramatic in this case.  On
the other hand, it appears that the application of the $\hat{L}^2$
constraint alone gives superior results to the application of both
$\hat{L}^2$ and $\hat{L}_z$ constraints in the cases of the
\textsuperscript{3}P $\to$ \textsuperscript{1}S transitions in carbon and
oxygen. We believe this behavior stems from an inconsistency in the
description of different $S$ and $L$ states in v2RDM methods in general.
For example, for linear chains of hydrogen atoms, we have
found\cite{FossoTande:2016:423} that large-$S$ states are more
well-constrained than low-$S$ states.  That effect, combined with an
apparent complementary effect regarding the relative description of
large-$L$ and small-$L$ states, results in estimates of the absolute
energies of the ${}^1$S states that are relatively poor, as compared to
estimates of the absolute energies of higher angular momentum states in
the same atoms (the absolute energies for all states considered here are
tabulated in the Supporting Information).  The application of $\hat{L}^2$
constraints alone (i.e., without constraints on $\langle \hat{L}_z
\rangle$) overstabilizes the \textsuperscript{3}P states, resulting is a
fortuitous cancellation of error in the description of the
\textsuperscript{3}P $\to$ \textsuperscript{1}S transitions in carbon and
oxygen. 

\begin{table}
    \centering

    \caption{\rm Energy differences (eV) between ground and excited spin
    and orbital angular momentum states calculated by at the v2RDM$^a$
    and full CI levels of theory. The lack of numerical data under the
    ``real'' heading indicates that the excited state in question is not
    accessible by v2RDM methods without considering angular momentum
    symmetry.}

    \label{TAB:LSTATES_ExcitationE} 
    \begin{tabular}{lccccccccccc}
        \hline
        \hline
        &atom	& ~~~ & transition                                      & ~~~ &  real    & ~~~ &   L$^2$               & ~~~ & L$_z$	            & ~~~ & full CI \\
\hline                                                                                                                                                  
        &Be	& ~~~ & \textsuperscript{1}S $\to$ \textsuperscript{3}P & ~~~ &  2.75    & ~~~ &   2.75                & ~~~ & 2.75                 & ~~~ & 2.75    \\
        &B	    & ~~~ & \textsuperscript{2}P $\to$ \textsuperscript{4}P & ~~~ &   3.56   & ~~~ &   3.56                & ~~~ & 3.52                 & ~~~ & 3.51    \\
        &C	    & ~~~ & \textsuperscript{3}P $\to$ \textsuperscript{1}D & ~~~ &  0.86    & ~~~ &   1.18                & ~~~ & 1.44                 & ~~~ & 1.49    \\
        &C	    & ~~~ & \textsuperscript{3}P $\to$ \textsuperscript{1}S & ~~~ &  --      & ~~~ &   2.80                & ~~~ & 2.68                 & ~~~ & 2.93    \\
        &C	    & ~~~ & \textsuperscript{3}P $\to$ \textsuperscript{5}S & ~~~ &  4.11    & ~~~ &   4.10                & ~~~ & 3.98                 & ~~~ & 3.93    \\
        &N	    & ~~~ & \textsuperscript{4}S $\to$ \textsuperscript{2}D & ~~~ &  1.75    & ~~~ &   2.07                & ~~~ & 2.57                 & ~~~ & 2.72    \\
        &N	    & ~~~ & \textsuperscript{4}S $\to$ \textsuperscript{2}P & ~~~ &  --      & ~~~ &   2.92                & ~~~ & 3.40$^b$             & ~~~ & 3.31    \\
        &N	    & ~~~ & \textsuperscript{4}S $\to$ \textsuperscript{4}P & ~~~ &  --      & ~~~ &   5.46                & ~~~ & 11.24                & ~~~ & 11.24   \\
        &O	    & ~~~ & \textsuperscript{3}P $\to$ \textsuperscript{1}D & ~~~ &  1.57    & ~~~ &   1.71                & ~~~ & 2.03                 & ~~~ & 2.14    \\
        &O	    & ~~~ & \textsuperscript{3}P $\to$ \textsuperscript{1}S & ~~~ &  --      & ~~~ &   4.28                & ~~~ & 3.82                 & ~~~ & 4.30    \\
        &F	    & ~~~ & \textsuperscript{2}P $\to$ \textsuperscript{4}P & ~~~ &  34.96   & ~~~ &  34.97                & ~~~ & 35.00$^b$            & ~~~ & 35.00   \\

                \hline
                \hline
    \end{tabular}
    \\

    ${}^a$ For values labeled as ``real,'' the specification of the
    spin angular momentum state is meaningful, while the specification
    of the orbital angular momentum state is not. \\
    $^b$ Loose convergence criteria were employed ($\epsilon_{\rm gap} <
    5.6\times10^{-4}$E$_{\rm h}$ and $\epsilon_{\rm error} <
    4.4\times10^{-6}$).\\

\end{table}

\begin{figure}[!htpb]

    \caption{The v2RDM energy (E$_{\rm h}$) for different $L_z$ projection
    states corresponding to the ${}^3$P and ${}^1$D terms of the carbon
    and oxygen atoms.}

    \label{FIG:LZ}

    \begin{center}
        \includegraphics[scale=1.0]{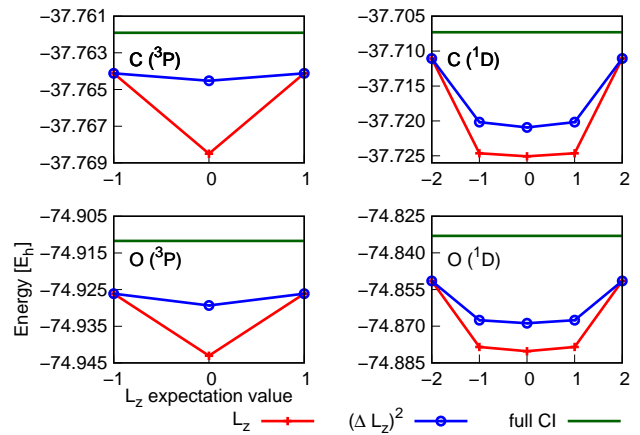}
    \end{center}
\end{figure}

To this point, all computations enforcing constraints on $\langle
\hat{L}_z \rangle$ considered only the maximal orbital projection state.
Here, we demonstrate that, for a given $L$-state, different orbital
angular momentum projections are not treated on equal footing by the v2RDM
approach.  Figure \ref{FIG:LZ} illustrates the energy for each $M_L$ state
within the manifold of states associated with the ${}^3$P and ${}^1$D
terms of the carbon and oxygen atoms.  For comparison, the horizontal
lines represent the corresponding full CI energies for each state.
Clearly, the v2RDM approach fails to recover the proper degeneracy of
different angular momentum projection states.  Rather, the v2RDM energy is
a convex function of the expectation value of $\hat{L}_z$, with the
maximal projection states giving the best lower-bound to the full CI
energy.  Similar observations were made by van Aggelen {\em et
al.},\cite{vanAggelen:2012:014110} regarding the treatment of spin
projection states within v2RDM theory.  The consideration of $\langle
\hat{L}_z \rangle = 0$ constraint does not improve the quality of the
v2RDM results over the case in which a real-valued algorithm is applied;
this result is not too surprising, since any purely real-valued 1-RDM
satisfies this constraint.  What is more interesting is that forcing the
variance $(\Delta L_z)^2$ to vanish substantially improves the quality of the
non-maximal orbital angular momentum projections, most dramatically so for
the $\langle \hat{L}_z \rangle = 0$ state; such a constraint could be
applied within a real-valued v2RDM optimization.  On the other hand,
variance constraints do not appear to improve the quality of the maximal
orbital angular momentum projection states.  Again, this behavior is
similar to that observed in Ref.  \citenum{vanAggelen:2012:014110} for
spin projection states.  In that work, the application of pure-state and
ensemble spin conditions yielded comparable results for maximal spin
projection states.

\subsection{Linear molecular systems}

\label{SEC:RESULTS_LINEAR}

Unlike the Hamiltonian for atomic systems, the Hamiltonian for linear
molecular systems does not commute with $\hat{L}^2$, so, in this case, the
only good orbital angular momentum quantum number is  $\Lambda = \langle
\hat{L}_z \rangle$, the projection of the orbital angular momentum on the
internuclear axis (which we have chosen to be aligned in the
$z$-direction).  The results presented above for atomic systems suggest
that orbital angular momentum projection constraints may play a similarly
important role in the v2RDM-based description of states with non-zero
$\Lambda$ (e.g., $\Pi$, $\Delta$, $\Phi$, etc.  states).  Hence, in this
Section, we explore the utility of constraints on $\hat{L}_z$ and $(\Delta
L_z)^2$ in linear molecular systems, beginning with a simple question: at the
v2RDM level of theory, is the ground state of molecular oxygen a singlet
or a triplet?

Table \ref{TAB:ST_GAP} illustrates the energy gap between the ${}^3\Sigma$
and ${}^1\Delta$ states of molecular oxygen, as computed at the v2RDM,
full CI, and MRCISD+Q levels of theory, in various basis sets.  Here, a
positive value for the gap indicates that the triplet is lower in energy.
Note that values labeled as ``real'' were generated without the
consideration of orbital angular momentum constraints, so the orbital
angular momentum is technically unspecified in these cases.  In a minimal
(STO-3G) basis, such a real-valued v2RDM computation predicts a
triplet/singlet gap of 0.914 eV, which is in reasonable agreement with
that from full CI (1.042 eV).  However, the v2RDM result is surprisingly
sensitive to the size of the basis set; in a 3-21G basis, the
triplet/singlet gap reduces to 0.424 eV, and, in a cc-pVDZ basis, the
singlet is actually predicted to be {\em lower} in energy than the triplet
by almost 0.2 eV.  Table \ref{TAB:ST_GAP} also provides results from
complex-valued v2RDM computations in which we have placed constraints on
the expectation value and variance of $\hat{L}_z$, where $\Lambda=0$ for
the triplet state ($^3\Sigma$) and $\Lambda=2$ for the singlet state
($^1\Delta$).  The application of orbital angular momentum constraints
significantly improves the v2RDM results, in all basis sets.  In
particular, $\hat{L}_z$ and $(\Delta\hat{L}_z)^2$ constraints remedy the
qualitative failure of the v2RDM approach within the cc-pVDZ basis. In
this case, the predicted triplet/singlet gaps are 0.924 eV and 0.940 eV,
respectively, which are both in reasonable agreement with the value of
1.049 eV predicted by MRCISD+Q.

\begin{table}

    \caption{The relative energies (eV) of the ${}^3\Sigma$ and
    ${}^1\Delta$ states of molecular oxygen,${}^a$ with an inter-atomic
    distance of 1.208 \r{A}.}

     \label{TAB:ST_GAP} 
    \begin{tabular}{lcccccc}
        \hline
        \hline
                                      & ~~~~ & STO-3G      & ~~~~ &  3-21G  & ~~~~ & cc-pVDZ \\
                \hline                                                      
                    MRCISD+Q          & ~~~~ & 1.042${}^b$ & ~~~~ &  1.113  & ~~~~ &  1.049  \\	
                    real              & ~~~~ & 0.914       & ~~~~ &  0.424  & ~~~~ & -0.196  \\	
                    L$_{z}$           & ~~~~ & 1.031       & ~~~~ &  1.132  & ~~~~ &  0.924  \\	
                    $(\Delta$L$_z)^2$ & ~~~~ & 1.037       & ~~~~ &  1.162  & ~~~~ &  0.940  \\	
                \hline
                \hline
    \end{tabular}
    \\

    {\small

    ${}^a$ For values labeled as ``real,'' the specification of the
    spin angular momentum state is meaningful, while the specification
    of the orbital angular momentum state is not. \\
    
    ${}^b$ This value was obtained from the full CI.

	}

\end{table}

\begin{figure*}[!htpb]

    \caption{The relative energies (eV) of the spin and orbital angular
    momentum states of molecular oxygen described by the (a) STO-3G, (b)
    D95V, and (c) cc-pVDZ basis sets. All energies are given relative to
    that of the ${}^3\Sigma$ state.}

    \label{FIG:O2_ENERGY_DIAGRAM}
    \includegraphics[scale=1.0]{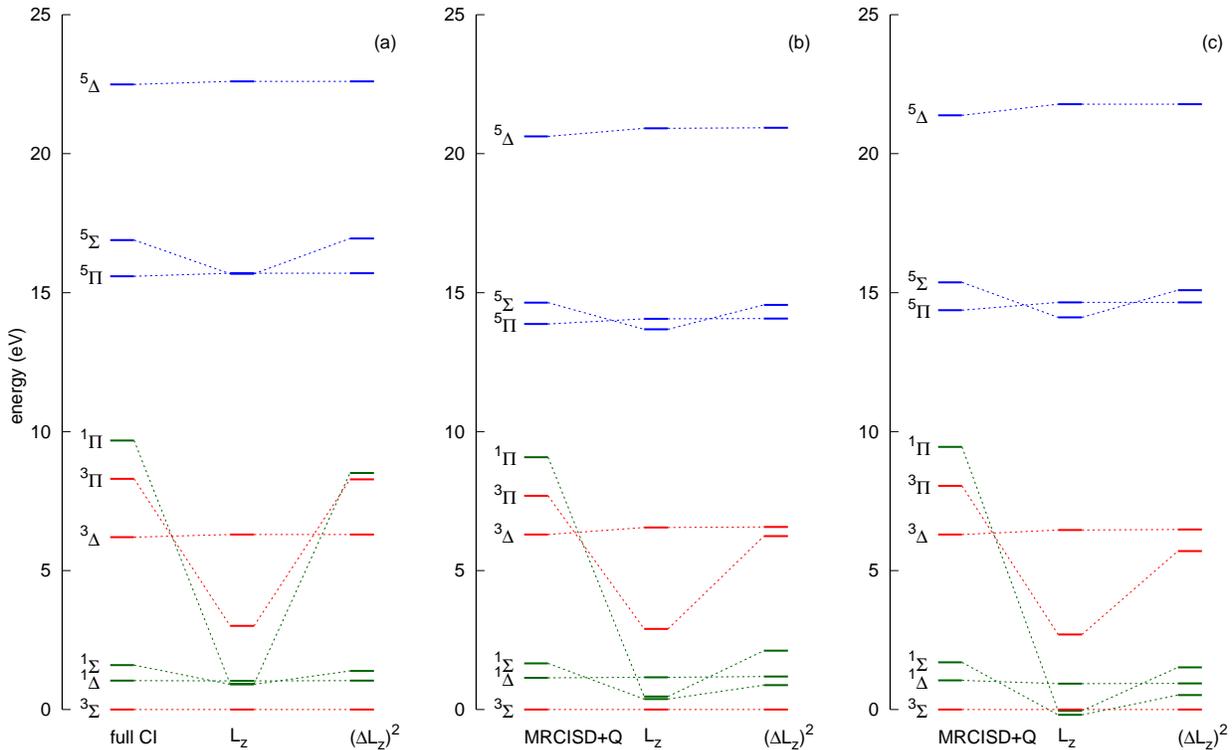}

\end{figure*}

In the cc-pVDZ basis set, the imposition of orbital angular momentum
constraints is clearly important for obtaining the correct ordering of the
spin angular momentum states of molecular oxygen.  However, these
constraints cannot guarantee the correct ordering of orbital angular
momentum states within a given spin manifold; this trend is evident in
energy diagrams depicted in  Fig. \ref{FIG:O2_ENERGY_DIAGRAM}.  In these
diagrams, the energy levels in all cases are shifted such that the energy
of the ${}^3\Sigma$ state is zero.  In a minimal basis set [Fig.
\ref{FIG:O2_ENERGY_DIAGRAM}(a)], the full CI, v2RDM [L$_z$], and v2RDM
[($\Delta$L$_z$)$^2$] approaches all predict that the ${}^3\Sigma$ is the
ground state. When constraining only the expectation value of $\hat{L}_z$,
the v2RDM approach incorrectly predicts that the three singlet states
considered are nearly degenerate, and the energy of the ${}^1\Pi$ state in
particular is severely underestimated.  Further, the energies of the
${}^5\Sigma$ and ${}^3\Pi$ states are far too low.  With variance
constraints, the v2RDM approach recovers the correct ordering for all spin
and orbital angular momentum states, but the spacing between the ground
and ${}^1\Pi$ state is still underestimated by more than 1 eV.  In the
D95V and cc-pVDZ basis sets [Figs. \ref{FIG:O2_ENERGY_DIAGRAM}(b) and
\ref{FIG:O2_ENERGY_DIAGRAM}(c), respectively], we observe similar dramatic
failures of the v2RDM approach (with constraints on the expectation value
of $\hat{L}_z$) to yield the correct state orderings, relative to the
orderings obtained from MR-CISD+Q.  In the cc-pVDZ basis in particular,
constraints on the expectation value of $\hat{L}_z$ alone are insufficient
to yield the correct ground state; the ${}^1\Sigma$ and ${}^1\Pi$ states
are {\em both} predicted to lie below the ${}^3\Sigma$ state. Fortunately,
the application of variance constraints leads to the correct prediction
that the ground state of molecular oxygen is a triplet.  Nonetheless, in
both the D95V and cc-pVDZ basis sets, the singlet and triplet states are
not ordered correctly amongst themselves; energies of the ${}^1\Pi$,
${}^1\Sigma$, and ${}^3\Pi$ states are all severely underestimated.  The
relative energies of all of the states considered in Fig.
\ref{FIG:O2_ENERGY_DIAGRAM} are tabulated in the Supporting Information.

\begin{figure}[!htpb]

    \caption{Dissociation curves for molecular oxygen, calculated within
    the D95V basis set.  The v2RDM computations enforced constraints on
    the expectation value and variance of $\hat{L}_z$. }

    \label{FIG:O2_DISSOCIATION}
    \begin{center}
        \includegraphics[scale=1.0]{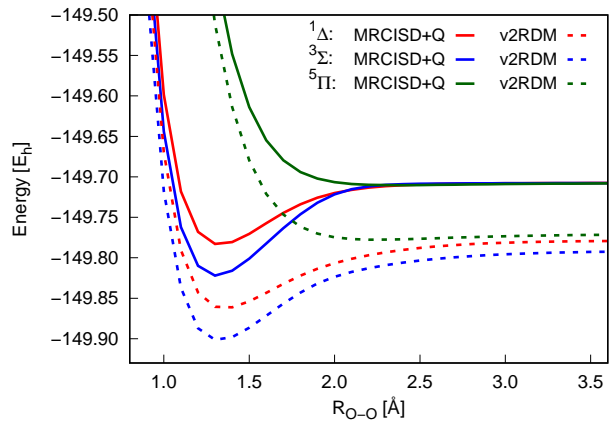}
    \end{center}
\end{figure}

Figure \ref{FIG:O2_DISSOCIATION} provides dissociation curves for the
${}^3\Sigma$, ${}^1\Delta$, and ${}^5\Pi$ states of O$_2$, as computed at
the v2RDM and MRCISD+Q levels of theory, within the D95V basis set.  Here,
the v2RDM curves were generated under orbital angular momentum constraints
($\langle \hat{L}_z \rangle = \Lambda$ and $(\Delta L_z)^2 = 0$), as well
as the spin angular momentum constraints outlined in Sec.
\ref{SEC:THEORY} for the maximal spin projection states.  As observed in
Table \ref{TAB:ST_GAP}, the ${}^3\Sigma$ / ${}^1\Delta$ energy gap is
well-predicted by the v2RDM approach at the equilibrium geometry, but the
overall shapes of the v2RDM-derived curves are not particularly accurate.
It is clear that the v2RDM approach suffers from some serious
deficiencies, particularly in the limit of dissociation.  The
${}^3\Sigma$, ${}^1\Delta$, and ${}^5\Pi$ curves should all share the same
energy at dissociation, but they do not, regardless of the imposition of
angular momentum constraints.

\begin{figure}[!htpb]

    \caption{The energy of molecular oxygen ($E_{\rm h}$), as described by
    the D95V basis set, at an O--O distance of 5 \AA.  The v2RDM
    computations enforced constraints on the expectation value of
    $\hat{L}_z$ or both the expectation value and variance of
    $\hat{L}_z$.}

    \label{FIG:O2_9DEGENERATESTATES}

    \begin{center}
        \includegraphics[scale=1.0]{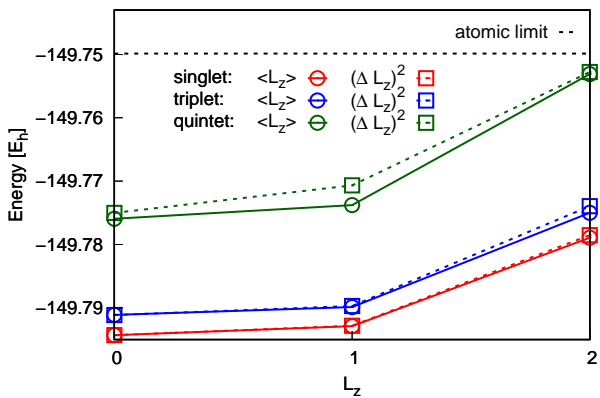}
    \end{center}

\end{figure}

The lack of degeneracy of the ${}^3\Sigma$, ${}^1\Delta$, and ${}^5\Pi$
states in the limit of dissociation is similar to the behavior observed in
Ref.  \citenum{vanAggelen:2012:014110}.  Those authors focused mainly on
the lack of degeneracy among different $M_S$ states, and it is clear from
that work that the maximal spin-projection states are the most well
constrained, in general (i.e., these states have the highest energies).
Here, we can draw similar conclusions regarding the orbital angular
momentum projections.  In the limit of dissociation, the ground state
should have an energy equal to twice that of a single oxygen atom in its
ground state (${}^3$P).  Two such atoms could couple to form nine states
with $S$ = 0, 1, or 2 and $\Lambda$ = 0, 1, or 2, all of which should be
degenerate at large O--O bond distances.  Figure
\ref{FIG:O2_9DEGENERATESTATES} illustrates the energy of these nine states
at an O--O bond length of 5.0 \AA; in all cases, the spin-projection state
is chosen to be the maximal one.  The dashed line represents twice the
energy of an isolated oxygen atom in the ${}^3$P state, as described by
the v2RDM method (constraining the maximal spin and orbital angular
momentum projection states, but not the expectation value of $\hat{L}^2$).
We can draw two conclusions from these data. First, for a given spin
state, higher orbital angular momentum projection states are more well
constrained. Second, for a given orbital angular momentum projection
state, the highest-multiplicity state is the most well constrained.
Indeed, the highest energy is obtained for the ${}^5\Delta$ state; the
size consistency error ($E_{\rm O_2}$ - 2 $E_{\rm O}$) is only 2.9 m$E_{\rm h}$ in this case.

\begin{figure}[h]

    \caption{Dissociation curves for the $^1\Sigma$ and $^1\Delta$ states
    of molecular carbon, calculated using the 6-31G* basis set. The v2RDM
    computations enforced constraints on the expectation value of
    $\hat{L}_z$, and the full CI results were taken from Ref.
    \citenum{Abrams:2004:9211}.}

    \label{FIG:C2_DISSOCIATION}

    \begin{center}
        \includegraphics[scale=1.0]{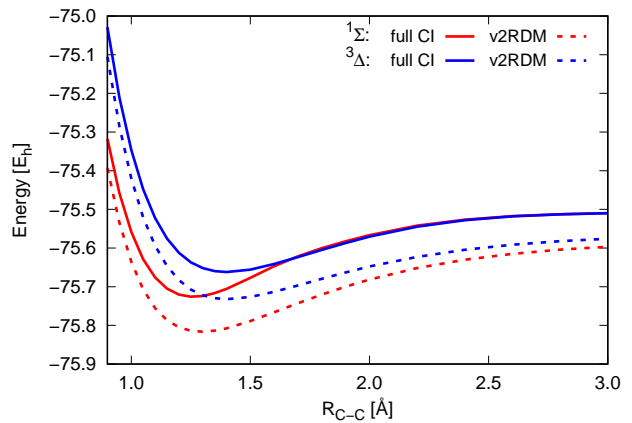}
    \end{center}

\end{figure}

Lastly, we consider dissociation curves for the $^1\Delta$ and $^1\Sigma$
states of another linear molecular system, C$_2$.  It is well known that a
proper description of these states requires a sophisticated treatment of
electron correlation
effects,\cite{Abrams:2004:9211,Booth:2011:084104,Mazziotti:2007:052502}
and, in the absence of orbital angular momentum constraints, v2RDM methods
can only describe whichever state lies lower in energy.  What is more
problematic is that, because the potential energy curves for the
$^1\Sigma$ and $^1\Delta$ states should cross, a real-valued v2RDM
computation may yield RDMs for different electronic states at different
C--C bond lengths.  Figure \ref{FIG:C2_DISSOCIATION} illustrates v2RDM and
full CI potential energy curves for C$_2$ computed within the 6-31G* basis
set.  Full CI results were taken from Ref. \citenum{Abrams:2004:9211}.
The application of orbital angular momentum constraints facilitates the
description of both states via the v2RDM approach, and, near the
equilibrium geometry for the ground state, we observe reasonable
splittings between the ground and excited states.  At a C--C  bond length
of 1.25 \AA, full CI predicts that the $^1\Delta$ state lies 2.43 eV above
the $^1\Sigma$ state, while the v2RDM approach predicts that these states
are separated by 2.90 eV.  The relative overstabilization of the
$^1\Sigma$ state is consistent with our observation that, for a given spin
state, higher orbital angular momentum projection states are more
well-constrained.  Unfortunately, the v2RDM method exhibits two
qualitative failures for this system.  First, it predicts that the
${}^1\Sigma$ state is the ground state for all C--C bond lengths; that is,
the potential energy cures for the two states are predicted to never
cross.  Second, as was observed above for molecular oxygen, the two
electronic states considered here do not share the same dissociation
limit.

\section{Conclusions}

\label{SEC:CONCLUSIONS}

In systems with well-defined orbital angular momentum symmetry, the
application of orbital angular momentum constraints facilitates the direct
variational determination of 2-RDMs for multiple electronic states.
Moreover, without such considerations, the v2RDM approach cannot
qualitatively describe states with non-zero $z$-projection of the orbital
angular momentum, even if the state in question is the lowest-energy state
of a given spin symmetry.  Indeed, we demonstrated that, in the absence of
orbital angular momentum constraints, the v2RDM approach incorrectly
predicts that the ground state of molecular oxygen (described by the
cc-pVDZ basis set) is a singlet.  The application of appropriate
constraints, which necessitates the consideration of complex-valued RDMs,
recovers the correct spin-state ordering.

The v2RDM energy appears to be a convex function of the expectation value
of $\hat{L}_z$, and, for a given magnitude of the orbital angular
momentum, maximal orbital angular momentum projection states are the most
well-constrained. This result reveals a qualitative failure of v2RDM
methods: they do not to recover the correct degeneracy for different
$L$/$M_L$ states, at least when the RDMs satisfy the ensemble
$N$-representability conditions considered in this work.  This behavior
suggests that the conclusions of Ref.  \citenum{vanAggelen:2012:014110}
regarding the description of different spin projection states apply to
angular momentum projection states in general.  Presumably, should one
consider the direct optimization of 2-RDMs corresponding to different {\em
total} angular momentum states, similarly incorrect behavior would emerge.

\vspace{0.3cm}
{\bf Acknowledgments}
\vspace{0.3cm}

This work was supported as part of the Center for Actinide Science and
Technology (CAST), an Energy Frontier Research Center funded by the U.S.
Department of Energy, Office of Science, Basic Energy Sciences under Award
No. DE-SC0016568.

\bibliography{angular_momentum}

\end{document}